# Periodically Correlated Time Series and the Variable Bandpass Periodic Block Bootstrap

Edward Valachovic
evalachovic@albany.edu
Department of Epidemiology and Biostatistics, School of Public Health,
University at Albany, State University of New York, One University Place,
Rensselaer, NY 12144

**Abstract**
This research introduces a novel approach to resampling periodically correlated (PC) time series using bandpass filters for frequency separation called the Variable Bandpass Periodic Block Bootstrap (VBPBB) and then examines the significant advantages of this new method. While bootstrapping allows estimation of a statistic's sampling distribution by resampling the original data with replacement, and block bootstrapping is a model-free resampling strategy for correlated time series data, both fail to preserve correlations in PC time series. Existing extensions of the block bootstrap help preserve the correlation structures of PC processes but suffer from flaws and inefficiencies. Analyses of time series data containing cyclic, seasonal, or PC principal components often seen in annual, daily, or other cyclostationary processes benefit from separating these components. The VBPBB uses bandpass filters to separate a PC component from interference such as noise at other uncorrelated frequencies. A simulation study is presented, demonstrating near universal improvements obtained from the VBPBB when compared with prior block bootstrapping methods for periodically correlated time series.

## 1. Introduction

**1.1 Background**
To better understand the characteristics of a set of data and the population the set of data represents, bootstrapping, or resampling with replacement from a given set of data, is useful to estimate the sampling distribution of statistics such as means and variances. Random sampling with replacement results in independent draws that form a resample of the same length as the original data set and was first detailed by Efron.[1] In time series analysis successive data points, or observations, are ordered in time. More on time series including definitions, notation, and examples can be found in Wei.[2] The ordering can also be in space, space-time, or another combination of dimensions. Generally described as spatio-temporal data, but without loss of generality, the data can still be referred to as time series and the metric that data is ordered in can be units of time. Often, ordered data points may be correlated with prior observations in the time series. Consequently, independently sampling data points from the dataset to form a new ordered resample will not replicate any correlations between a given and prior data points within the original time series. The correlation between successive time series data points is destroyed. Special bootstrapping schemes are needed to prevent the destruction of correlation structures within time series data.

Block bootstrapping is a class of methods which attempt to replicate and preserve correlations in spatio-temporal data. Block bootstrapping often involves a general strategy of splitting the time series into blocks and then randomly sampling the blocks to form the resamples. An example of this is the Moving Block Bootstrap first introduced by Kunsch.[3] This can help replicate the correlation structure between given data points and, at least up to a limited number, some prior observations. Clearly, correlations between data points separated by more than the length of the bootstrapped blocks will have correlation structures destroyed.

Many fields across the natural and social sciences involve spatio-temporal data driven by periodic, also called cyclic or seasonal, factors or components. Periodic components with a given period, $p$, exhibit strong correlations between data points that are $kp$ time points removed, or lagged, where $k$ is an integer multiple. Such a component of a time series is periodically correlated (PC). In general, block bootstrapping has difficulty reproducing the correlation structure of PC time series. As previously described, block bootstrapping with any block lengths less than the given period, $p$, will sever these correlations. Additionally, bootstrapping with block lengths greater than the given period, $p$, will preserve the correlation within each block; however, there is no reason randomly sampled adjacent blocks will retain synchronicity in the periodic correlation structure when forming the resample. In either case, there is nothing to force the first data point of a sampled block to be the next step in each cycle of period $p$ that follows the last data point of the prior sampled block. For this reason, many block bootstrapping strategies are not well suited for PC time series.

**1.2 Existing Methods and Challenge**
A block bootstrapping strategy better suited for PC time series with a PC component of period, $p$, is one that considers bootstrapping blocks whose length in some way accounts for the period $p$. A seasonal block bootstrap was proposed by Politis where the block length is restricted to be a multiple of the period $p$.[4] Regardless of whichever step in the cycle a block begins, all other blocks in the resample will likewise start and end with a common step in the cycle of period $p$. This strategy will preserve correlation structures between data points that are p points removed from each other in time series data. Some other block bootstrapping methods for PC time series similar to this include those presented by Chan et al. and Dudek et al.[5,6]

In many real-world situations, time series will be a function of the accumulated influence of multiple components, such as a noise or random error component in addition to the PC component. A common source for these problems stems from the specific field (meteorology, environmental sciences, physics, economics) as well as measurement error and instrument precision problem common to all fields. While the block bootstrap methods described above, which can collectively be referred to as periodic block bootstrap (PBB) methods, attempt to preserve the PC correlation structure, these strategies also bootstrap any additional components including the noise component. Unsurprisingly, even modest levels of noise can widen bootstrapped confidence intervals for measures such as the PC component periodic mean more than the supposed confidence interval level, a result confirmed by this work's simulations. Figure 1 illustrates the PBB bootstrap method of choosing block size to preserve the correlation structure of the PC component while simultaneously resampling all other components, represented as a noisy signal, in the PC time series.

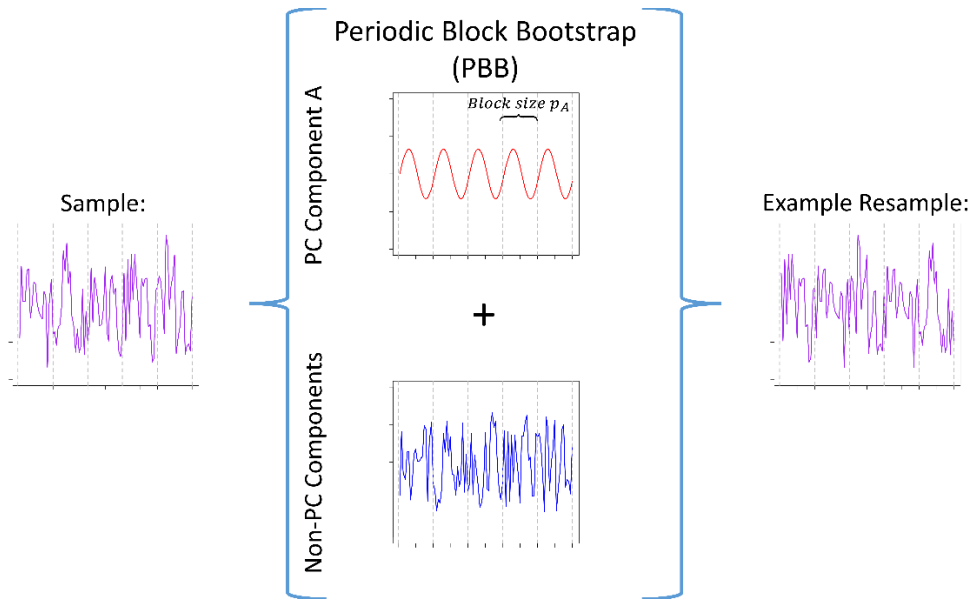

**Figure 1.** Illustration of a PBB resampling a PC time series composed of a PC component A with period $p_A$, and non-PC components. The PBB resamples blocks of size $p_A$ from all components, replicating the PC structure of component A while simultaneously resampling from the non-PC components.

### 1.3 VBPBB Approach
The objective is to bootstrap the PC time series component alone to preserve the correlation structures of the PC component while not bootstrapping other unrelated components that do not share the given period of the PC component. A solution is inspired by the periodogram, which is a spectral density or frequency domain representation of a time series and is described in Wei.[2] Components operating at different frequencies, or the reciprocal of period, may sum in the time domain but have no correlation between them. In this way, different frequencies operate independently. Filtering a PC time series according to its spectral density by separating a PC component by its frequency, would create a new PC time series composed primarily of the component PC of interest, while preserving the PC component correlation structure. The new component PC time series could then be block bootstrapped with the appropriate block size to preserve that correlation structure. This approach, or Variable Bandpass Periodic Block Bootstrapping (VBPBB), should preserve the desired correlation structures within a PC time series while greatly reducing the bootstrapping of uncorrelated components. The VBPBB strategy is illustrated in Figure 2.

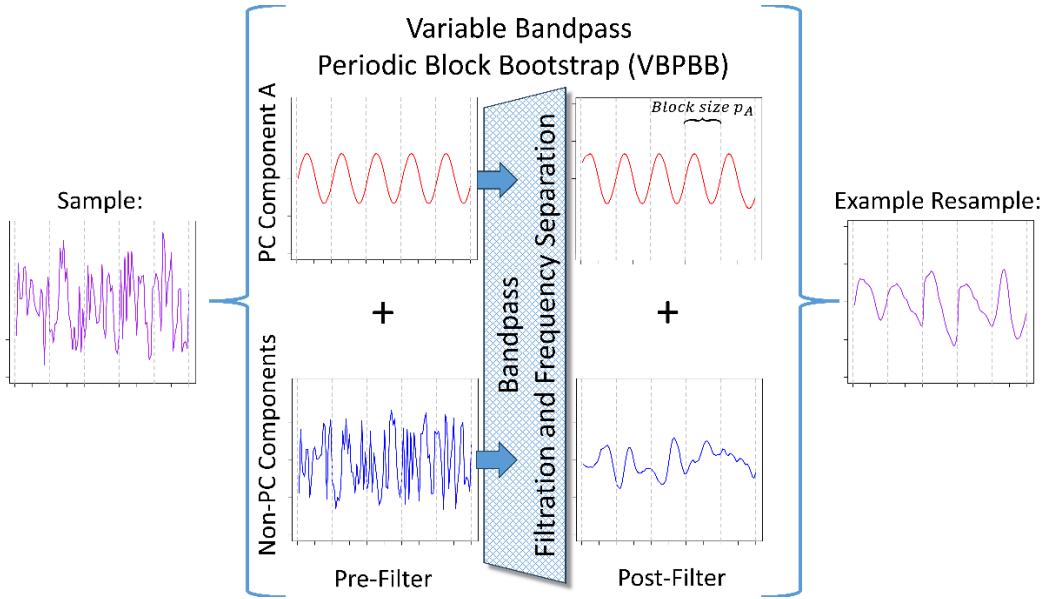

**Figure 2.** Illustration of VBPBB resampling a PC time series composed of a PC component A with period $p_A$, and non-PC components. The VBPBB first bandpass filters to pass PC component A, while reducing non-PC components. VBPBB then resamples blocks of size $p_A$ from the bandpass filtered components, replicating the PC structure of component A while limiting replication of the non-PC components.

There are several potential advantages to VBPBB with PC time series. This approach is not tied to a specific periodic block bootstrapping technique, so each of the PBB methods previously mentioned could benefit from frequency separation. VBPBB can be accomplished with a variety of filters that may best suit different situations. This work proposes using a class of low pass filters, the Kolmogorov-Zurbenko (KZ) filters, and their band pass extensions, but others may be considered. Finally, through the exclusion of uncorrelated components, VBPBB has the potential to produce more accurate estimates of the sampling distribution that are also more robust against noise and other forms of interference.

In VBPBB it is necessary to separate and filter the PC time series by a portion of the spectral density; what is passed through the filter contains the PC component frequency. This can be accomplished by applying a bandpass filter to pass a narrow band of frequencies around the corresponding frequency of the PC component, while frequencies outside this band, the bandstop, are attenuated. Consequently,

PBB methods are essentially a special case of VBPBB, when the bandpass filter is made trivially wide and passes the entire spectrum of frequencies.

A variety of filters, alone or in combination, can produce the desired result. One such filter is the KZ filter and its extensions. These filters are described in Zurbenko.[7] KZ filters are a class of low-pass filters, but their extensions include bandpass filters, and in combination with difference filters, are a flexible way to produce low, high, and bandpass filters with fine control over filter cut-off frequencies, or boundary between the passband and stopband. KZ filters are iterations of a simple moving average. These filters are well suited for computing processes that include resampling algorithms. Furthermore, the parameters of the filter provide a clear and direct explanation relating to the desired problem.

KZ filters and their extensions can separate portions of the frequency domain to exclude interfering frequencies as detailed in Yang and Zurbenko.[8] These filters have natural uses to isolate frequencies in a variety of fields such as the environmental sciences, meteorology, economics, public health, and climatology. Examples include investigating ozone in Tsakiri and Zurbenko, air quality in Kang et al., and global temperature in Ming and Zurbenko.[9,10,11] Examples also include atmospherics in Zurbenko and Potrzeba, climate in Zurbenko and Cyr, and more recently diabetes in Arndorfer and Zurbenko.[12,13,14] The bandpass separated component can then independently be investigated to reveal important details about patterns and processes often hidden within the original data, as well as associations with possible factors operating at similar spatio-temporal scales. This is the idea behind spatio-temporal analysis by frequency separation (STAFS). A novel example of this applied to skin cancer is extensively explored in Valachovic and Zurbenko.[15,16] Frequency separation is used to identify hidden PC components in a skin cancer time series and then to perform multivariate analysis on these component factors in Valachovic and Zurbenko.[17] VBPBB adapts the use of frequency separation prior to analysis and proposes its adaptation to a bootstrap design for PC time series. These examples highlight the use of KZ filters to smooth data, reduce random variation, interpolate missing observations, and more specifically, to separate and filter portions of the frequency domain prior to analysis. These features make KZ filters and their extensions ideal for use in VBPBB.

## 2. Methods

### 2.1 Bandpass Filter

The Kolmogorov-Zurbenko (KZ) filter is the iteration of a simple central moving average defined in Zurbenko.[7] It is a filter with two parameters $m$, and $k$. The parameter $m$, a positive odd integer, is the filter window length and the parameter $k$, a positive integer, is the number of iterations. KZ filters are low pass filters that strongly attenuate signals of frequency $1/m$ and higher while passing lower frequencies, smoothing the time series. Equation 1 is a KZ filter applied to a random process $\{X(t): t \in \mathbb{Z}\}$ with $m$ time points, and $k$ iterations:

$$KZ_{m,k}(X(t)) = \sum_{u=-k(m-1)/2}^{k(m-1)/2} \frac{a_u^{m,k}}{m^k} X(t+u), \tag{1}$$

where coefficients $a_u^{m,k}$ are the polynomial coefficients from:

$$\sum_{r=0}^{k(m-1)} z^r a_{r-k(m-1)/2}^{m,k} = (1 + z + \cdots + z^{m-1})^k$$

One advantage of the KZ filter is the computational ease with which statistical software can apply it in an iterated form. As an application of k iterations of a moving average filter of *m* time points, the KZ filter can be produced according to the algorithm in Equation 2:

$$KZ_{m,1}(X(t)) = \sum_{u=-(m-1)/2}^{(m-1)/2} \frac{a_u^{m,1}}{m^1} X(t+u) = \frac{1}{m} \sum_{u=-(m-1)/2}^{(m-1)/2} X(t+u)$$

$$KZ_{m,2}(X(t)) = \frac{1}{m} \sum_{u=-k(m-1)/2}^{k(m-1)/2} KZ_{m,1}(X(t+u))$$

$$\vdots$$

$$KZ_{m,k}(X(t)) = \frac{1}{m} \sum_{u=-k(m-1)/2}^{k(m-1)/2} KZ_{m,k-1}(X(t+u)) \tag{2}$$

The transfer function is the linear mapping describing how input frequencies are transferred to outputs. The energy transfer function is the square of the transfer function and as such is symmetric about zero. The energy transfer function of the KZ filter at frequency $\lambda$ is seen Equation 3. This shows how with only a few iterations, a KZ filter, strongly attenuates signals of frequency 1/*m* and higher while passing lower frequencies:

$$|B(\lambda)|^2 = \left(\frac{\sin(\pi m \lambda)}{m \sin(\pi \lambda)}\right)^{2k} \tag{3}$$

The cut-off frequency is a limit or boundary at which the energy transferred through a filter is suppressed or diminished rather than allowed to pass through. The point where output power is $\alpha \in (0,1)$ times that of the input can be used as the boundary. It is common to use $\alpha = 1/2$ or the half power point, a power ratio in $10 * log_{10}$ of -3 decibels units. The cut-off frequency $\lambda_0$ for the transfer function for a KZ filter is provided in Equation 4.

$$\lambda_0 \approx \frac{\sqrt{6}}{\pi} \sqrt{\frac{1-\left(\frac{1}{2}\right)^{\frac{1}{2k}}}{m^2-\left(\frac{1}{2}\right)^{\frac{1}{2k}}}} \tag{4}$$

Where the KZ filter is a low pass filter, strongly filtering signals of a frequency at or above the frequency equivalent to $1/m$, and may be useful for isolating low frequencies, the related Kolmogorov-Zurbenko Fourier Transform (KZFT) filter is a band pass filter. Equation 6 is a KZFT filter applied to a random process $\{X(t): t \in T\}$ having three parameters: $m$ time points, $k$ iterations, and shifted center at a frequency $\nu$:

$$KZFT_{m,k,\nu}(X(t)) = \sum_{u=-k(m-1)/2}^{k(m-1)/2} \frac{a_u^{m,k}}{m^k} e^{-i2m\nu u} X(t+u), \tag{5}$$

where the coefficients $a_u^{m,k}$ are the polynomial coefficients from:

$$\sum_{r=0}^{k(m-1)} z^r a_{r-k(m-1)/2}^{m,k} = (1 + z + \cdots + z^{m-1})^k$$

Where the KZ filter is symmetric around zero, the KZFT is a symmetric band pass filter around frequency $\nu$. Practical use of the KZFT filter is similar to the KZ filter since it can be produced in statistical software. The energy transfer function of the KZFT filter at a frequency $\lambda$ with parameters $m$, $k$, and $\nu$ is seen Equation 6.

$$|B(\lambda - \nu)|^2 = \left(\frac{\sin(\pi m(\lambda - \nu))}{m \sin(\pi(\lambda - \nu))}\right)^{2k} \tag{6}$$

The cut-off frequency $\lambda_0$ for the transfer function for a KZFT filter is provided in Equation 7.

$$|\lambda_0 - \nu| \approx \frac{\sqrt{6}}{\pi} \sqrt{\frac{1-\left(\frac{1}{2}\right)^{\frac{1}{2k}}}{m^2-\left(\frac{1}{2}\right)^{\frac{1}{2k}}}} \tag{7}$$

For these filters, the cut-off frequency boundaries then become useful to determine the region of the spectra passed verses those stopped or suppressed.

**2.2 VBPBB Parameters**

KZFT filters are used prior to block bootstrapping methods to separate the PC component in PC time series. To better understand the construction of the KZFT filters used to separate the PC component prior to block bootstrapping, we consider each filter parameter and their interpretation. Recall the KZFT filter has three parameters, $m$, $k$, and $\nu$. The parameter $\nu$ is the easiest to interpret since it is the frequency around which the filter will be centered, and will be set to 1/period, the reciprocal of the period of the PC component of interest. The KZFT filter is symmetric around frequency $\nu$.

The KZ filter parameter $m$ defines the width of the moving average filter window. It can be interpreted as defining the endpoints of the filter width, where the time series energy is essentially eliminated. The effect of varying the $KZ$ filter parameter $m$ is illustrated in Figure 3.

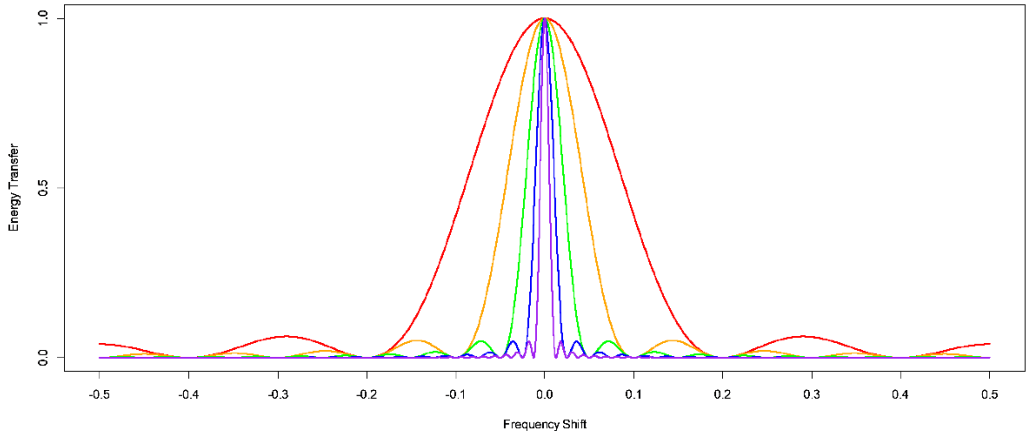

**Figure 3**. The energy transfer function at different frequency shifts for a KZ filter centered at frequency zero, or equivalently a KZFT filter centered at frequency $\nu$, with parameters $k = 1$ and $m = 5$ in red, $m = 11$ in orange, $m = 21$ in green, $m = 41$ in blue, and $m = 81$ in purple.

The KZ filter parameter $k$ defines the number of iterations of the moving average filter performed. It can be interpreted as defining the taper of the filter and it can move the filter cut-off even though it does not change the filter endpoints where the time series energy is completely attenuated. The effect of varying the $KZ$ filter parameter $k$, for a constant fixed $m$, is illustrated in Figure 4.

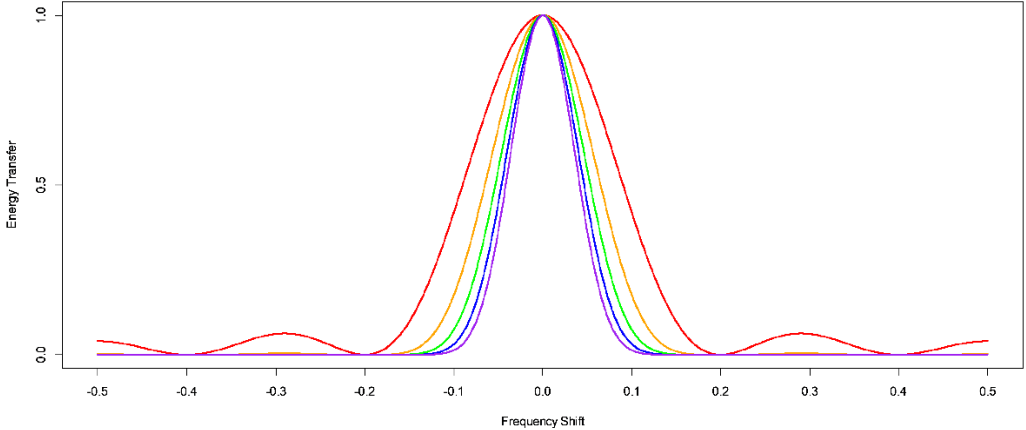

**Figure 4**. Energy transfer functions at different frequency shifts for a KZ filter centered at frequency zero, or equivalently a KZFT filter centered at frequency ν, parameters $m = 5$ and $k = 1$ in red, $k = 2$ in orange, $k = 3$ in green, $k = 4$ in blue, and $k = 5$ in purple.

Understanding the control provided by the parameter choice of KZFT filters enables these tools to take a PC time series and filter out other frequencies lying outside of a narrow band around a specific PC component frequency. The PC component time series is now primarily composed of the PC correlation structure. Provided there is a sufficiently lengthy spatio-temporal PC time series dataset relative to the parameter choices to support implementing the filters, KZFT filters offer the flexibility to filter any PC component.

Finally, after PC component separation, the VBPBB approach next block bootstraps the PC component time series, unlike PBB which bootstraps the original PC time series. While it is possible to use any of the PBB methods with VBPBB outlined earlier that preserve a PC component of a fixed period or frequency, this work utilizes a block bootstrapping method similar to the General Seasonal Block Bootstrap (GSBB) of Dudek et al.[6] For a given period, $p$, a time series of $n$ observations is block bootstrapped by creating $p$ exclusive and exhaustive subsets, each composed of one of each of the first $p$ observations, and the integer multiple observations of that observation. Each resample is formed by randomly selecting from the $p$ subsets in sequence, repeated until n observations are selected. VBPBB then repeats this process for a large number, $B$, of resamples. Statistics, such as the periodic mean are calculated for each resample, providing an estimate of the sampling distribution for that statistic.

### 3. Simulations

This work tests VBPBB performance applied to PC time series for consistency and robustness through simulations conducted under scenarios where the assumed conditions and settings are comparable to real world PC time series data analysis. Each simulation produces a PC time series dataset for use with the PBB approach and an identical dataset for the new VBPBB approach that uses bandpass filters to recover the PC component prior to using the identical block bootstrapping method as the PBB approach.

**3.1 Simulation Methods**
Analysis is performed in R version 4.1.1 statistical software and uses the KZFT function in the KZA package, see Close and Zurbenko for more detail, with simulated datasets of a time series.[18,19] The simulations are performed across a range of scenarios of PC component periods, or frequencies, and under different levels of noise. All time series are constructed with 1000-time units. First, a sine wave signal with a fixed period, p, where the time coordinate determines the phase of the sine wave, is simulated with an amplitude of one producing the PC component. Next, random variation is introduced by generating equal length vectors of elements randomly selected from a standard normal distribution with variance scaled to give the desired signal to noise ratio. These random variations are then summed with the PC component. The same simulated dataset is then used for the two approaches, first with PBB by blocking to preserve the PC correlation structure for the PC period by fixing the block size to p, but without separating the PC component frequency. Next, on the identical dataset, VBPBB separates the PC component using KZFT filters and then block bootstraps the component time series according to the described strategy, with fixed block size $p$. In the VBPBB approach, KZFT filters are centered above the PC component frequencies, setting $v = 1/p$.

This simulation study examines many scenarios with different combinations of PC component period and signal to noise ratio. The KZFT parameters $m$ and $k$ could be adjusted for more optimal performance in each of these scenarios. However, to simplify this simulation study presented here, the KZFT parameters are initially fixed at $m = 11$ and $k = 1$ across all scenarios. These parameter choices may not be optimal, as measured by the metrics that will be used to compare the VBPBB and PBB approaches. Further questions of optimal parameter choice for different scenarios and different objectives are best left for future work. In each scenario, $B = 1000$ bootstrap resamples are recorded and 95% bootstrapped confidence intervals for the PC component periodic mean are produced from the 0.975 and 0.025 quantiles. To test consistency, in each scenario the production of the $B = 1000$ bootstrap resamples and accompanying confidence intervals are repeated 1000 times for each approach and results aggregated. Finally, to test robustness, these simulations are performed within the scenarios of different PC periods including {10, 25, 50, 100, 250} to span a wide range of possible frequencies, and at a signal-to-noise ratio including {1:2, 1:5, 1:10} to represent a wide range of interference from sources, such as noise. Additionally, to test robustness in scenarios at combinations of the five given periods and at the same levels of noise,

VBPBB and PBB are used to attempt to block bootstrap a null PC component, or when no variation at a given PC period is present. In total, across the PC component scenarios and the null PC component, including the three levels of signal to noise ratio, five different periods, 1000 repetitions and $B = 1000$ bootstrap resamples in each repetition, this simulation study represents 30 thousand applications (30 million simulated resamples) of VBPBB and PBB each.

Figure 5 illustrates an example simulation of a PC component, a noise component, and their summation which produces a PC time series which is what would be typically observed for analysis. The figure then shows the bandpass filter reconstruction of the PC component.

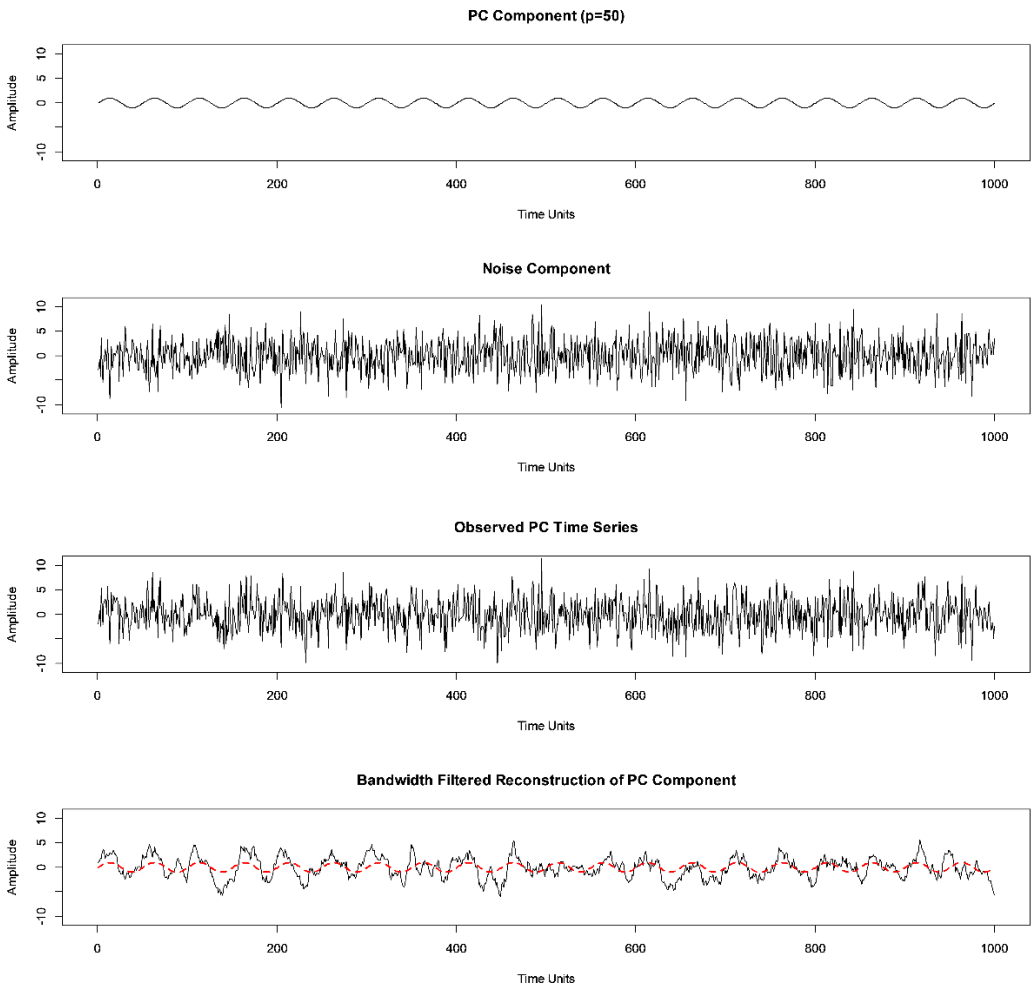

**Figure 5.** (From top) Example of simulated PC component with p=50, noise component, observed PC time series summing components, and a bandpass filtered reconstruction of the PC component with original PC component in red.

### 3.2 Assessment of the VBPBB

To assess the performance of PBB of the PC time series verses that of the VBPBB, we look at how well each method has replicated the correlation structure of the PC time series. At each PC period and signal-to-noise ratio, and for each of the 1000 repetitions, the size of a 95% confidence interval (CI) for the periodic mean is created. The median size of these 1000 repetitions is calculated for the PBB and VBPBB approaches. In theory, we expect the CI size to be smaller for VBPBB. Additionally, for each repetition the proportion of the PC component lying within the 95% CI is calculated, and then the median proportion is found among the 1000 repetitions.

Last, the median of the periodic means for the 1000 repetitions is calculated for the two approaches and correlated with the original PC component used in the simulation to see how well the component correlation structure has been preserved. Apart from the correlation comparison, the same metrics are used to compare approaches on simulated data where there is no PC component present, to see if the methods are robust to misspecification.

Figure 6 illustrates an example of one of the simulated repetitions for a PC component at period $p = 50$ with signal to noise ratio 1:10. Here, the PBB 95% CI in red averages 1.96 times larger than the VBPBB 95% CI in blue. Neither approach has any portion of the PC component fall outside of the respective CI. The square of correlation between the PC component and the median periodic mean from VBPBB is 0.89. The square of correlation between the PC component and the median periodic mean from PBB is 0.33. Interpreting the square of correlation as the proportion of variation explained, this implies the median periodic mean from the VBPBB approach explains 56 percent greater variability in the PC component than that from the PBB approach. Therefore, in this example, CIs with VBPBB are approximately half the size with significantly greater fidelity to the PC correlation structure with no loss in ability to capture the true PC component.

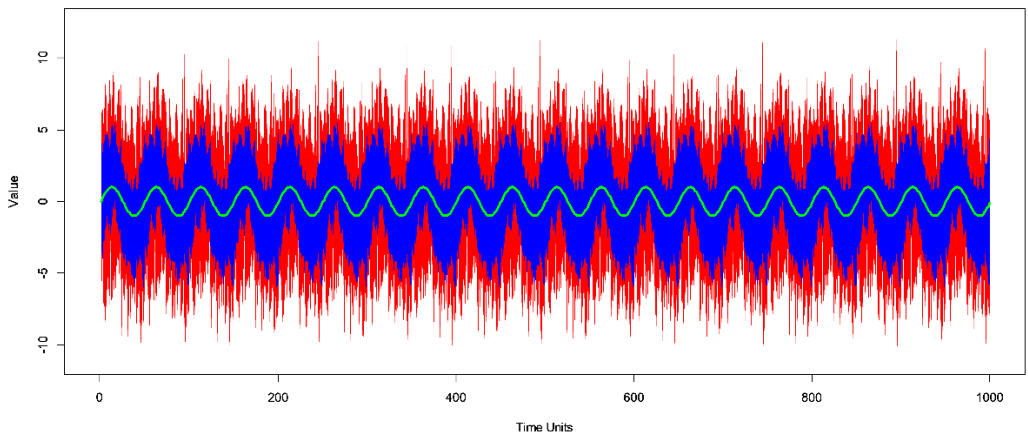

**Figure 6.** Example of a simulated PC component with $p = 50$ in green, 95% CI for the periodic mean from PBB in red and the 95% CI for the periodic mean from VBPBB in blue.

## 4. Results

The relative performance of the new VBPBB approach in comparison to that of PBB is ascertained through the metrics and results of the simulations testing these approaches. The robustness of these approaches is addressed by simulating results across scenarios of different PC component periodicities and under different conditions of interference by noise. This includes the simulation of a null or no PC component. The consistency of these approaches is addressed by repeating each scenario simulation 1000 times for every combination of PC period and noise and then aggregating results.

For each combination of PC component period and noise, the Table 1 provides the median (across the 1000 scenario repetitions) ratio of 95% CI size comparing PBB to VBPBB. The value represents the multiplicative factor, or how many times bigger, the 95% CI size is of the PBB approach as compared to the VBPBB approach. The larger the CI, the worse the performance of the approach. A ratio of one indicates PBB and VBPBB performed similarly, while the larger the ratio above one the better the performance of the VBPBB approach. Across the scenarios of the simulation study design, ratios range 1.58 to 2.24, demonstrating widespread superior performance by VBPBB. This implies CIs produced by PBB are typically 58% to 124% larger than those from VBPBB.

**Table 1**. Median ratio of 95% CI sizes for the periodic means comparing PBB to VBPBB.

| Period (Frequency) | Signal to Noise Ratio | | |
|---|---|---|---|
| | 1:2 | 1:5 | 1:10 |
| 10 (0.10) | 2.24 | 2.24 | 2.21 |
| 25 (0.04) | 2.01 | 2.00 | 2.02 |
| 50 (0.02) | 1.70 | 1.69 | 1.70 |
| 100 (0.01) | 1.61 | 1.61 | 1.61 |
| 250 (0.04) | 1.58 | 1.60 | 1.60 |

For each combination of PC component period and noise, the Table 2 provides the median (across the 1000 scenario repetitions) difference in the square of correlation between the approaches' periodic means and the true PC component; for VBPBB, less that from PBB. The value represents the additional percentage of the PC component variance in the periodic mean explained by the VBPBB approach that is not explained by the PBB approach. The higher the square of correlation, or percent of variance explained, the better the approach performs in preserving the PC correlation structure. A difference of zero indicates PBB and VBPBB performed similarly, while the more positive the difference the better the performance of the VBPBB approach. Excluding the scenarios when the period is 10, differences range from 2.84 to 53.68, again supporting widespread better performance by VBPBB. This implies VBPBB explained approximately 3% to 54% more of the variability of the original PC component than from VBPBB. For the scenarios with the period of 10, PBB outperformed VBPBB when the KZFT filter parameter is $m = 11$. Recall, for the simulation study design, the VBPBB used KZFT filters with fixed parameters $m = 11$ and $k = 1$, and the result at a period of 10 demonstrates this selection was not necessarily ideal in all scenarios. This highlights of the need for flexibility of the parameters used for VBPBB previously mentioned. Indeed, the simulations were repeated for the scenarios with a period of 10, but with the KZFT parameter changed to $m = 5$ instead of $m = 11$ (asterisks on table). With that small change, VBPBB again outperformed PBB across the scenarios with differences ranging from 3.80 to 15.01. This change came with the trade-off of CI size for the periodic mean, but with only a modest reduction in the advantage of VBPBB. With the change, for scenarios where the period is 10, CI size for PBB was 57% to 59% larger than from VBPBB.

**Table 2**: Median difference of the square of correlation between the periodic means and the true PC component for VBPBB subtracting that from PBB.

| Period (Frequency) | Signal to Noise Ratio | | |
|---|---|---|---|
| | 1:2 | 1:5 | 1:10 |
| 10 (0.10) | 3.80* | 8.71* | 15.01* |
| 25 (0.04) | 2.84 | 16.33 | 29.35 |
| 50 (0.02) | 17.93 | 32.83 | 45.61 |
| 100 (0.01) | 33.02 | 48.20 | 53.68 |
| 250 (0.04) | 48.89 | 51.81 | 46.02 |

*Note*: * Asterisks indicate for these simulation scenarios VBPBB used parameters $m = 5$ and $k = 1$ instead of $m = 10$ and $k = 1$.

For each combination of PC component period and noise, the Table 3 provides the median (across the 1000 scenario repetitions) difference in percentage of the PC component time series that falls outside a 95% CI for the periodic means, VBPBB less that from PBB. The lower the percentage of the PC component time series falling outside of bootstrapped 95% CIs for the periodic mean, the better. For a procedure producing 95% CIs, it would be anticipated, on average, approximately 5% of the periodic mean should fall outside of the interval. A difference of zero indicates PBB and VBPBB performed similarly, while the larger the difference the better the performance of the PBB approach. Given the much larger CI sizes of PBB it is anticipated to perform better here, however, VBPBB performance similar to PBB would imply the much smaller CI size using VBPBB comes at little to no cost. Except for select scenarios when the period is 250 and 100, differences were all less than 0.05, providing widespread evidence CI accuracy did not suffer under VBPBB compared to PBB. Of note are the excepted scenarios when the period is 250 or 100, where VBPBB underperformed. Again, these select results illustrate the need for flexibility of the parameters used for VBPBB. Recall, the simulation study design for VBPBB used KZFT filters with fixed parameters $m = 11$ and $k = 1$. For these scenarios, when simulations are repeated with the KZFT parameter changed to $m = 5$ (asterisks on table), performance of VBPBB improved in each scenario with a difference of <0.05 compared to PBB. Here, this change came with a trade-off of CI size for the periodic mean, where CI size for PBB was now 11% to 13% larger than that from VBPBB.

**Table 3**: Median difference in percentage of the PC component time series falling outside a 95% CI for the periodic means for VBPBB subtracting that from PBB.

| Period (Frequency) | Signal to Noise Ratio | | |
|---|---|---|---|
| | 1:2 | 1:5 | 1:10 |
| 10 (0.10) | <0.01 | <0.01 | <0.01 |
| 25 (0.04) | <0.01 | <0.01 | <0.01 |
| 50 (0.02) | <0.01 | <0.01 | <0.01 |
| 100 (0.01) | <0.05* | <0.05 | <0.05 |
| 250 (0.04) | <0.05* | <0.05* | <0.05* |

*Note*: * Asterisks indicate for these simulation scenarios VBPBB used parameters $m = 5$ and $k = 1$ instead of $m = 10$ and $k = 1$.

Lastly, for each noise level when there is in fact no PC component present (the null case), the Table 4 provides the median (across the 1000 scenario repetitions) ratio of 95% CI size comparing PBB to VBPBB. The value represents the multiplicative factor, or how many times bigger, the 95% CI size is of the PBB approach as compared to the VBPBB approach. The larger the confidence interval, the worse the performance. A ratio of one indicates PBB and VBPBB performed similarly, while the larger the ratio above one, the better the performance of the VBPBB approach. Across the scenarios, with the null PC component simulation study design, ratios range 1.59 to 2.24, supporting widespread better performance by VBPBB. This implies CIs produced by PBB are typically 59% to 124% larger than those from VBPBB.

**Table 4**: Median ratio of 95% CI sizes for the periodic means comparing PBB to VBPBB.

| Period (Frequency) | Noise (Variance) | | |
|---|---|---|---|
| | 2 | 5 | 10 |
| 10 (0.10) | 2.24 | 2.24 | 2.21 |
| 25 (0.04) | 2.01 | 2.00 | 2.02 |
| 50 (0.02) | 1.70 | 1.69 | 1.70 |
| 100 (0.01) | 1.61 | 1.61 | 1.61 |
| 250 (0.04) | 1.59 | 1.60 | 1.60 |

Since there is no PC component in these scenarios, there is no square of correlation between the periodic means and the true null PC component to compare between VBPBB and PBB. It is however possible to compare the median difference in percentage of the null PC component time series that falls outside a 95% CI for the periodic means for VBPBB subtracting that from PBB. It was found in all scenarios the percentage of the true null PC component time series falling outside a 95% CI for the periodic means were nearly identical between VBPBB and PBB. The median difference in percentage of the null PC component time series falling outside a 95% CI for the periodic means for VBPBB subtracting that from PBB had all values less than <0.01. Therefore, a table of these values is not included.

### 5. Discussion

This work proposes the VBPBB as an improvement upon prior bootstrapping methods for the preservation of the correlation structure when resampling a PC time series. The results contained here strongly support the conclusion that even a rudimentary or inflexible application of the VBPBB approach produces almost uniformly superior results to other PBB methods for the preservation of the correlation structure and accurate estimation of the sampling distribution. The simulation study showed VBPBB robustness across a wide range of scenarios of PC component frequency and interference from noise, including PC component misspecification with little to no loss in accuracy or consistency. The theory and design of the VBPBB approach shows that PBB methods inadvertently but unnecessarily leaves other components unrelated to the PC component of interest such as noise in the PC time series resampling process, which VBPBB provides a means of mitigating. While the PBB approach does have some success preserving the PC component correlation structure, it is inefficient compared to VBPBB. For bootstrapping statistics such as the periodic mean of the PC component in these simulations, the CI size for PBB is generally larger, often significantly, than that produced by VBPBB. Furthermore, PBB simulated CI size was systematically too large for the stated CI percentage. In the few scenarios where VBPBB was less successful than PBB, adaptability of the VBPBB improved performance in comparison to PBB, and with only modest trade-offs in the metrics measuring success.

VBPBB has several limitations in its design in comparison to other PBB methods. As just discussed, VBPBB performance is tied to selection of parameters used for bandpass filtration. As the simulation study results show in this work, there is no one universally best set or subset of parameters, and optimality in the choice of parameters in different scenarios is an open question and a potential subject of further work. Also, the VBPBB filters the PC time series to pass a certain bandwidth around a PC component frequency through moving averages, which essentially smooth data. While it might reasonably be assumed that continually decreasing bandwidth would lead to continually improved performance, it is possible to over-smooth the time series, causing a decrease in performance. This is another limitation and emphasizes the need for additional work for guidance on optimality. Since VBPBB methods rely upon filtering the original PC time series with the application of moving averages, the observations at the ends of the time series will have the filter incompletely applied. Hence, all else being equal, VBPBB generally requires a greater quantity of observed data than that of PBB, another limitation. Still, VBPBB requires marginally more data than other methods, the additional demand is modest and linked only to the parameter choices, so VBPBB retains the strengths of a non-parametric method.

The improvements made by VBPBB extend beyond the study of PC time series in the context of resampling. In a wider impact, VBPBB should benefit time series analysis in general. In many real-world examples time series are a mixture of components, and analysis or modeling of non-PC components can benefit from the improved estimation of the characteristics of a PC component, even if it is not the component of interest. Still, with the listed limitations in mind, the primary strength of VBPBB is the improved preservation of the PC time series correlation structure as seen in the superior correlation of the reconstructed seasonal mean and the original PC component compared to the PBB approach in the simulation studies of this work. This translated into reducing bootstrapped CIs for statistics with improved or at least no loss of accuracy. An additional strength includes little to no loss of consistency with obtained results over many simulated repetitions. Also, VBPBB exhibits a wider robustness against real world factors such as PC period and interference from other components such as noise, including resiliency against misspecification of the PC component frequency. These strengths make VBPBB a practical, widely applicable, and often superior option for block bootstrapping periodically correlated time series.